\documentclass[]{spie}
\usepackage{amsmath,amsfonts,graphics}

\newcommand{\Prob}	{ {\rm P} }

\newcommand{\Cmu}	{ C_\mu }

\newcommand{\LocalPast}			{ L^{-}} 
\newcommand{\localpast}			{ l^{-}}
\newcommand{\localpastprime}		{ \lambda}

\newcommand{\LocalFuture}		{ {L}^{+}}

\newcommand{\LocalState}		{ S}
\newcommand{\localstate}		{ s}

\title{Quantifying Self-Organization in Cyclic Cellular Automata}
\author{Cosma Rohilla Shalizi\supit{1} and Kristina Lisa Shalizi\supit{2}
\skiplinehalf
\supit{1}Center for the Study of Complex Systems, University of Michigan, Ann Arbor, MI 48109 USA
\skiplinehalf
\supit{2}Statistics Department, University of Michigan, Ann Arbor, MI 48109 USA
}
\authorinfo{Address correspondence to CRS.\\ 
E-mail: $\{$cshalizi,kshalizi$\}$@umich.edu}

\begin{document}
\maketitle
\begin{abstract}
Cyclic cellular automata (CCA) are models of excitable media.  Started from
random initial conditions, they produce several different kinds of spatial
structure, depending on their control parameters.  We introduce new tools from
information theory that let us calculate the dynamical information content of
spatial random processes.  This complexity measure allows us to quantitatively
determine the rate of self-organization of these cellular automata, and
establish the relationship between parameter values and self-organization in
CCA.  The method is very general and can easily be applied to other cellular
automata or even digitized experimental data.
\end{abstract}
\keywords{Self-organization, cellular automata, excitable media, cyclic
  cellular automata, information theory, statistical complexity,
  spatio-temporal prediction, minimal sufficient statistics}

\section{Introduction}

The term ``self-organization'' was introduced in the 1940s \cite{Ashby-1947},
and become one of the leading concepts of nonlinear science, without, however,
ever having received a proper characterization.  The prevailing ``I know it
when I see it'' standard actually impedes scientific progress, since it
prevents our developing even the rudiments of a {\em theory} of
self-organization.  Thus some researchers state that ``self-organizing''
implies ``dissipative'' \cite{Nicolis-Prigogine-self-organization}, and others
claim to exhibit reversible systems that self-organize \cite{Raissa-DLA}, and
no one can even say if they are both talking about the same idea.  It would be
useful, therefore, to have a definition of self-organization which was
mathematically precise, so we could build formal theories around it, and
experimentally applicable, so that one could say whether or not something
self-organizes on the basis of empirical data.  Any such definition must be
tested against cases where intuition is clear, preferably cases where we know
exactly what is going on.

We believe we have such a definition of self-organization, and test it here
against cellular automata, specifically the so-called cyclic cellular automata.
We selected these systems for a number of reasons: their dynamics are
completely known and can easily be simulated exactly, they are decent
qualitative models of chemical waves, and there is already an analytical theory
of the patterns they form, against which we can check our results.

\section{Spiral Wave Formation in Cyclic Cellular Automata}

Cyclic cellular automata
\cite{Fisch-Gravner-Griffeath,Fisch-Gravner-Griffeath-threshold-range} (CCA)
are caricatures of pattern formation in chemical oscillators and other
excitable media \cite{Winfree-time-breaks-down}.  Each site in a square
two-dimensional lattice is in one of $\kappa$ colors.  A cell of color $k$ will
change its color to $k+1 \bmod \kappa$ if there are already at least $T$ cells
of that color in its neighborhood, i.e., within a distance $r$ of that cell.
Otherwise, the cell retains its current color.  (All cells update their color
in parallel.)

The CCA has three generic forms of long-term behavior, depending on the size of
the threshold relative to the range.  At high thresholds, the CCA forms
homogeneous blocks of solid colors, which are completely static --- so-called
fixation behavior.  At very low thresholds, the entire lattice eventually
oscillates periodically; sometimes the oscillation takes the form of large
spiral waves which grow to engulf the entire lattice.  There is an intermediate
range of thresholds where incoherent traveling waves form, propagate for a
while, and then disperse; this is called ``turbulence'', but whether it has any
connection to actual fluid turbulence is unknown.  With a range one ``Moore''
(box) neighborhood, the phenomenology is as follows
\cite{Fisch-Gravner-Griffeath-threshold-range}.  $T=1$ and $T=2$ are both
locally periodic, but $T=2$ produces spiral waves, while $T=1$ ``quenches''
incoherent local oscillations.  At $T=3$, one encounters ``turbulence,'' which
is actually meta-stable --- spiral waves can form and entrain the entire CA,
but this does not always happen on finite lattices, and in any case the
turbulent phase can persist for very long times.  Fixation occurs with $T \geq
4$.  Intuitively, all three phases of the CCA can be said to self-organize when
started from uniform noise.  Also intuitively, that is, by the ``I know it when
I see it'' standard, the fixation phase is less organized than turbulence,
which is less organized than spiral waves.  It is hard to say, by eye, whether
incoherent local oscillations are more or less organized than simple fixation.

\begin{figure}
\begin{center}
\resizebox{0.5\textwidth}{!}{\includegraphics*[4.45in,3.35in]{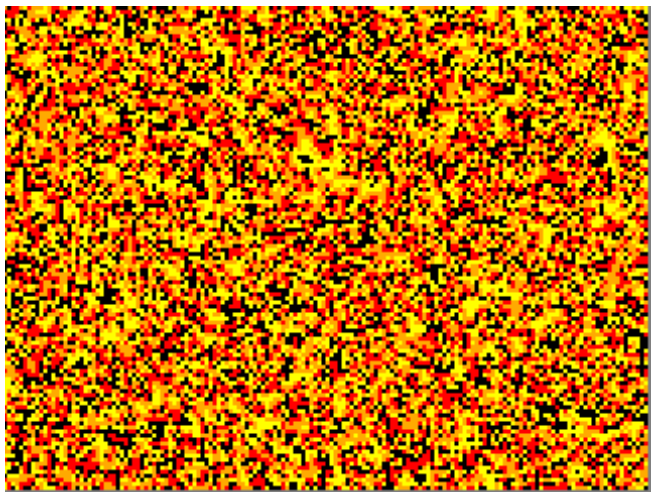}}
\end{center}
\caption{\label{fig:local-osc-shot} Local oscillations in the CCA $(T=1)$.  In
  this and the subsequent figures, the number of colors is fixed to 4, the rule
  uses a range-one box-shaped neighborhood, and the initial conditions are a
  uniform random mixture of colors.  These figures, but not our numerical
  results, were prepared using the MJCell program by Mirek W{\'o}jtowicz,
  \texttt{http://psoup.math.wisc.edu/mcell/}.  Observe the slight departure
  from uniform randomness, in the shape of connected twisting single-color
  bands.  The CA as a whole oscillates with period 4, each cell cycling through
  the colors.}
\end{figure}

\begin{figure}
\begin{center}
\resizebox{0.5\textwidth}{!}{\includegraphics*[4.45in,3.35in]{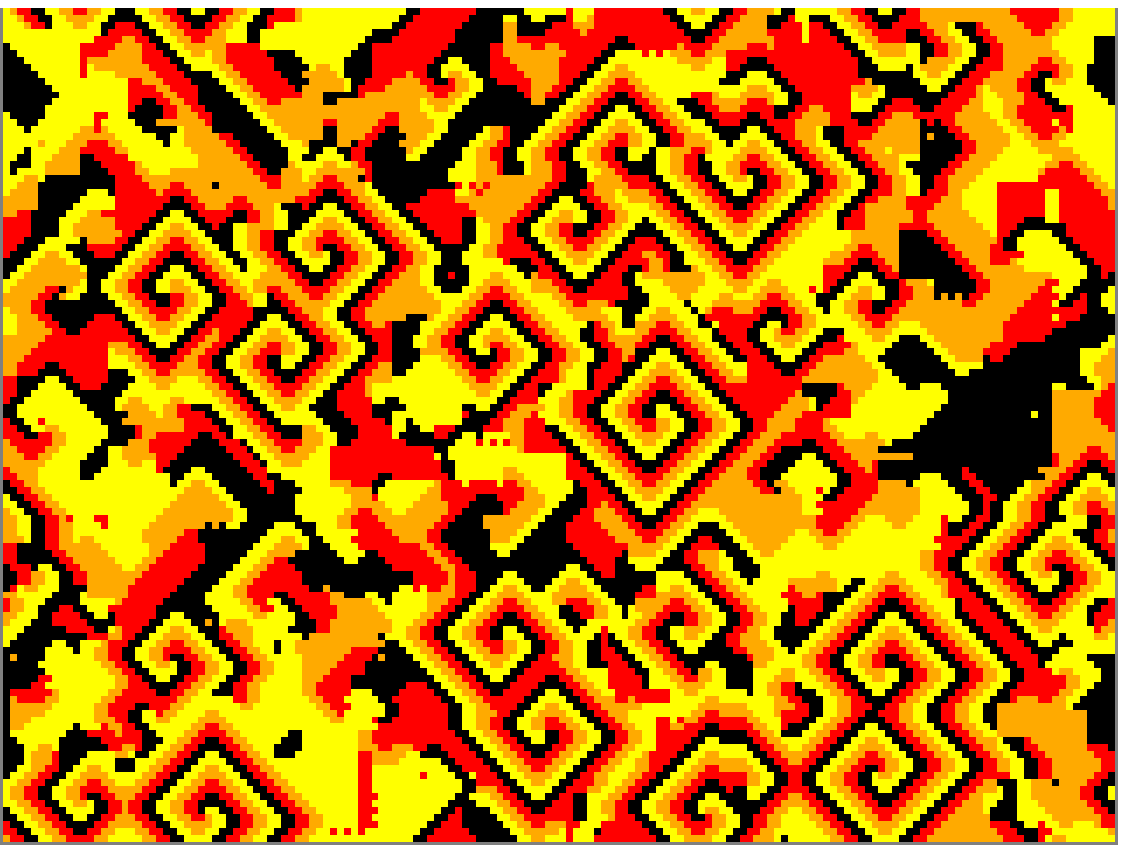}}\resizebox{0.5\textwidth}{!}{\includegraphics*[4.45in,3.35in]{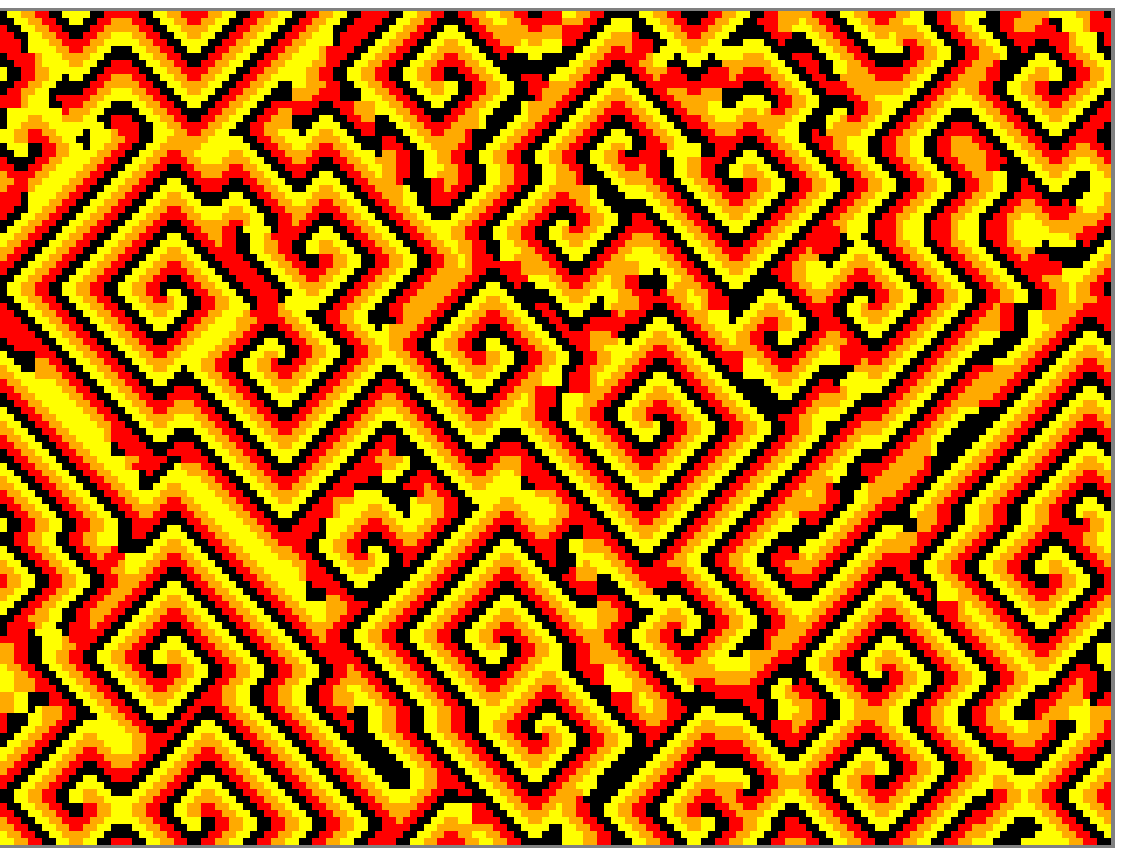}}
\end{center}
\caption{\label{fig:spiralwave-shot} Spiral waves $(T=2)$ in the cyclic CA.
  Left: formation of spirals out of background regions, $t = 45$.  Right: the
  same system later ($t = 200$), when the spirals have entrained the entire
  lattice.}
\end{figure}

\begin{figure}
\begin{center}
\resizebox{0.5\textwidth}{!}{\includegraphics*[4.45in,3.35in]{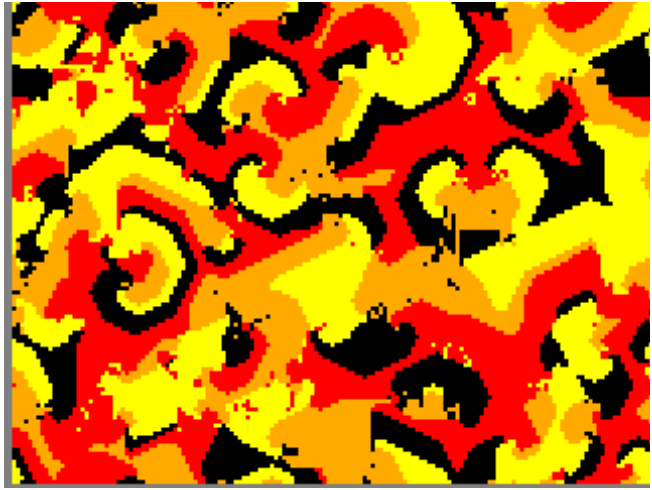}}
\end{center}
\caption{\label{fig:turbulence-shot} Typical view of the CCA ``turbulent''
  phase $(T=3)$.}
\end{figure}

\begin{figure}
\begin{center}
\resizebox{0.5\textwidth}{!}{\includegraphics*[4.45in,3.35in]{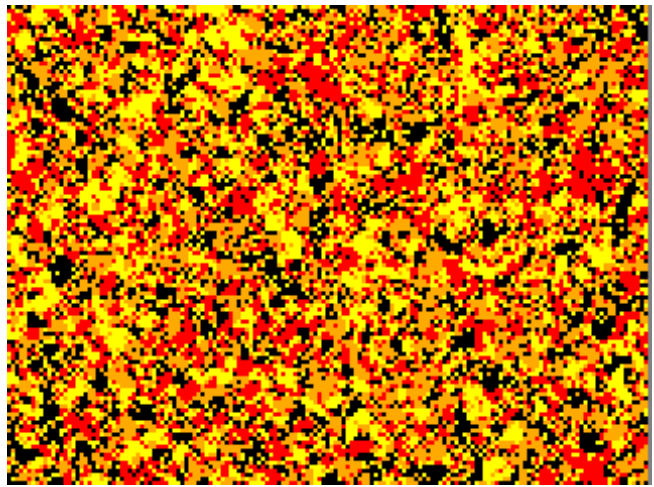}}
\end{center}
\caption{\label{fig:fixation-shot} Fixation in the CCA $(T=4)$.  Note that the
  configuration departs slightly from the random initial condition, owing to
  the growth of rectangular blocks of solid color.}
\end{figure}

\section{A Measure of Organization}

Few attempts have been made to measure self-organization quantitatively in
either model or real systems.  (See
Refs.\ \citenum{Gershenson-Heylighen-self-organizing,CRS-thesis} for reviews.)
There is a widespread sense, perhaps first articulated by Bennett
\cite{Bennett-dissipation,Bennett-1986,Bennett-how-and-why}, that
self-organization is the same as a spontaneous increase in complexity, leaving
us with the problem of measuring complexity.  The obvious candidate, from a
physical point of view, is thermodynamic entropy, and at least two studies have
claimed to measure self-organization by measuring spontaneous declines over
time in entropy \cite{Wolfram-stat-mech-CA,Klimontovich}.  Unfortunately
configurational or thermodynamic entropy is a most unsatisfying measure of
organization in complex systems
\cite{Medawar-general-evolution,Fox-energy-evolution,Badii-Politi,DPF-JPC-why,%
  Sewell-emergent-macrophysics}.

We shall expand on that last point, because it is not as well-appreciated as it
should be.  Thermodynamic entropy measures the degree of a system's
``mixedupedness'' (to use Gibbs's word), or how far it departs from being in a
pure state\cite{Sewell-emergent-macrophysics}.  For many of the systems treated
by statistical mechanics, pure states are more organized than impure ones, but
there is no logical connection.  Low-temperature spin systems may be in very
pure states which have no significant organization.  Organisms are essentially
never in pure states, and are highly mixed up at the molecular level, but are
the paradigmatic examples of organization.  In fact, there are many instances
of biological self-organization which are thermodynamically favored because
they {\em increase}
entropy\cite{Medawar-general-evolution,Fox-energy-evolution}.  Furthermore,
there are many different {\em kinds} of organization, and entropy ignores
all the distinctions and gradations between them\cite{DPF-JPC-why}.

So much for the idea that organization is simply ``unmixedupedness''.  Another
school of thought, going back to Kolmogorov\cite{Kolmogorov-three-approaches}
and Solomonoff \cite{Solomonoff} holds that complex phenomena are ones which do
not admit of descriptions which are both short and accurate.  The idea that
complexity should be identified with long minimal descriptions has been
implemented in many different ways \cite{Badii-Politi,Li-and-Vitanyi-1993}.
Many of these implementations remain in Kolmogorov's original framework of {\em
  exactly} describing a {\em particular} configuration, and so inherit the
feature that independent random configurations are hard to describe.  (For the
record of heads and tails from tossing a fair coin, the shortest algorithmic
description is usually the sequence itself.)  This is paradoxical from a
stochastic point of view, where independent variables are the most basic kind
available.  The paradox may be resolved by changing our goal, to {\em
  statistically} describing {\em ensembles} of
configurations.\cite{Rissanen-SCiSI} (The heads-and-tails sequence is
described as ``independent Bernoulli trials with parameter $p=0.5$'', or,
colloquially, ``something you'd get by tossing a lot of coins''.)  Within this
general framework, the most satisfying version of the description-length idea
defines the complexity of a process as the minimal amount of information about
its state needed for maximally accurate prediction.  Grassberger
\cite{Grassberger-1986} first proposed this idea, and Crutchfield and Young
\cite{Inferring-stat-compl} gave operational definitions of ``maximally
accurate prediction'' and ``state''.  For a full exposition of the resulting
theory, as it applies to classical stochastic processes, see
Ref.\ \citenum{CMPPSS}.

The Crutchfield-Young ``statistical complexity'', $\Cmu$, of a dynamical
process is the Shannon entropy (information content) of the minimal sufficient
statistic for predicting the process's future \cite{CMPPSS}.  In thermodynamic
settings, this is just equal to the amount of information a full set of
macrovariables contains about the microscopic state of the system
\cite{CRS-thesis,What-is-a-macrostate}.\footnote{The fact that $\Cmu$ and the
  thermodynamic entropy are both ``entropies'' is a merely verbal, or at best
  algebraic, resemblance; their physical meanings and theoretical functions are
  entirely different.}  For spatially extended systems, which are of interest
here, it is more appropriate to look at the density of statistical complexity,
using a refinement of the theory that handles local quantities, which we
briefly sketch here, following Ref.\ \citenum{CRS-thesis}.\footnote{Further
  details will be presented in a paper one of us (CRS) is writing with Robert
  Haslinger.}

\subsection{Minimal Local Sufficient Statistics}

Take a random field varying over space and time $X(\vec{r},t)$, where there is
some maximum speed $c$ at which information can propagate.  The {\em past light
  cone} of the space-time point $(\vec{r}, t)$ consists of all points which
could influence $X(\vec{r},t)$, i.e., all points $(\vec{s}, u)$ such that $u <
t$ and $|\vec{s} - \vec{r}| \leq c(t-u)$.  Similarly, the {\em future light
  cone} of $(\vec{r},t)$ is the set of all points which could be influenced by
what happens there.  Write $\LocalPast(\vec{r},t)$ for the configuration of the
random field in the past light cone, and $\LocalFuture(\vec{r},t)$ for the
configuration in the future light cone.  (We will suppress the arguments except
when comparing cones at different locations.)  There is a certain distribution
over future light-cone configurations conditional on the configuration in the
past.  The full notation for this would be $\Prob(\LocalFuture|\LocalPast =
\localpast)$, but we shall abbreviate it as $\Prob(\LocalFuture|\localpast)$.

Any function $\eta$ of $\LocalPast$ defines a {\em local statistic} or {\em
  local effective state}.  It can be thought of as summarizing the influence of
all the space-time points which could affect what happens at $(\vec{r},t)$.
Such local statistics should tell us something about ``what comes next,'' which
is $\LocalFuture$.  In fact, we can use information theory to quantify how
informative different statistics are.

The Shannon entropy or information content of a random variable $Z$ is
\begin{eqnarray}
H[Z] & \equiv & -\sum_{z}{\Prob(Z=z)\log_2{\Prob(Z=z)}}\\
& = & -\left<\log_2{\Prob(Z)}\right> ~,
\end{eqnarray}
which is the average number of bits needed to encode the value of $Z$.
Equivalently, $H[Z]$ indicates how much uncertainty an ideal statistician would
have about $Z$.  The conditional entropy of $Z$ given $Y$,
\begin{eqnarray}
H[Z|Y] & \equiv & -\sum_{y}{\Prob(Y=y)\sum_{z}{\Prob(Z=z|Y=y)\log_2{\Prob(Z=z|Y=y)}}}\\
& = & -\left<\log_2{\Prob(Z|Y)}\right> ~,
\end{eqnarray}
is the average number of bits needed to encode $Z$ if one knows $Y$.  Hence the
reduction in uncertainty, or description length, due to a knowledge of $Y$ is
\begin{eqnarray}
I[Z;Y] & \equiv & H[Z] - H[Z|Y]\\
& = & \left<\log_2{\frac{\Prob(Z,Y)}{\Prob(Z)\Prob(Y)}}\right> ~,
\end{eqnarray}
called the {\em mutual information} shared by $Z$ and $Y$.

For our case, we want to consider the information a local statistic $\eta$
conveys about the future, which is $I[\LocalFuture;\eta(\LocalPast)]$.  A
statistic is {\em sufficient} if it is as informative as
possible\cite{Kullback-info-theory-and-stats}.  In this case, a statistic
$\eta$ is sufficient if and only if $I[\LocalFuture;\eta(\LocalPast)] =
I[\LocalFuture;\LocalPast]$, i.e., if it retains all the predictive information
in the complete past.  This in turn is equivalent to requiring that
$\Prob(\LocalFuture|\eta(\localpast)) = \Prob(\LocalFuture|\localpast)$, for
all past configurations.  Put slightly differently, given a sufficient
statistic, one does not need to remember the data from which one computed it.

All sufficient statistics have the same predictive ability, but they are not
equal in the resources they need to make that prediction.  In particular, if
one is using the local statistic $\eta$ to make predictions, one must describe
or encode $\eta$, which takes $H[\eta(\LocalPast)]$ bits.  If knowing the value
of one statistic, say $\eta_1$, lets us determine the value of another,
$\eta_2$, then intuitively $\eta_2$ is a more concise summary, and in fact
$H[\eta_1(\LocalPast)] \geq H[\eta_2(\LocalPast)]$.  A {\em minimal local
  sufficient statistic}\cite{Kullback-info-theory-and-stats} is one whose value
can be determined from any other sufficient statistic.  Unsurprisingly, minimal
local sufficient statistics minimize the entropy $H[\eta(\LocalPast)]$.  How
can we construct a minimal statistic?

Take any two past light-cone configurations, $\localpast_1$ and $\localpast_2$.
Each has some conditional distribution over future light-cone configurations,
$\Prob(\LocalFuture|\localpast_1)$ and $\Prob(\LocalFuture|\localpast_2)$
respectively.  Say that the two past configurations are equivalent,
$\localpast_1 \sim \localpast_2$ if those conditional distributions are equal.
Then for each past configuration $\localpast$, there is a set of configurations
which are equivalent to it, which we may write $[\localpast]$.  Finally,
consider the function which maps past configurations to their equivalence
classes under the relation $\sim$:
\begin{eqnarray}
\epsilon(\localpast) & \equiv & [\localpast]\\
& = & \left\{\localpastprime\left|\Prob(\LocalFuture|\localpastprime) = \Prob(\LocalFuture|\localpast)\right.\right\}
\end{eqnarray}
Clearly, $\Prob(\LocalFuture|\epsilon(\localpast)) =
\Prob(\LocalFuture|\localpast)$, and so $I[\LocalFuture;\epsilon(\LocalPast)] =
I[\LocalFuture;\LocalPast]$, making $\epsilon$ a sufficient statistic.  The
equivalence classes, the values $\epsilon$ can take, are called the (local)
{\em causal states}\cite{Inferring-stat-compl,CMPPSS}.  Each causal state
corresponds to a distinct conditional distribution for the contents of the
future light-cone.  We write the causal state at $(\vec{r},t)$ as
$\LocalState(\vec{r},t)$.

We saw already that $\Prob(\LocalFuture|\localpast) =
\Prob(\LocalFuture|\eta(\localpast))$, for any sufficient statistic $\eta$.  So
if $\eta(\localpast_1) = \eta(\localpast_2)$, then
$\Prob(\LocalFuture|\localpast_1) = \Prob(\LocalFuture|\localpast_2)$, and the
two pasts belong to the same causal state.  Hence, if we know the value of
$\eta(\localpast)$, we can determine the value of $\epsilon(\localpast)$.  But
this means that $\epsilon$ is a minimal sufficient statistic.  Moreover, one
can show\cite{CRS-thesis} that $\epsilon$ is the {\em unique} minimal
sufficient statistic, in the sense that any other minimal statistic is just a
relabeling of the same set of states.

Because $\epsilon$ is a minimal sufficient statistic, $H[\epsilon(\LocalPast)]
\leq H[\eta(\LocalPast)]$, for any other sufficient statistic $\eta$.  This
being the case, we are entitled to speak, in an objective manner, about the
minimal amount of information needed to predict the system, or, as we may also
think of it, how much information the system retains from its past.  This
quantity, $H[\LocalState]$ is a characteristic of the system, and not (just) of
a particular class of model.  We therefore write $\Cmu \equiv H[\LocalState]$,
and call this the {\em statistical complexity density}; we will frequently drop
the ``density''.  $\Cmu$ is the amount of information about the past light-cone
which the system's dynamics make relevant to the future.

We now propose to define self-organization thus: a system has self-organized
between time $t_1$ and time $t_2$ if (1) $\Cmu(t_1) < \Cmu(t_2)$, and (2) not
all of the increase in complexity is due to outside intervention.  Clearly, if
the system is not being manipulated from the outside at all, (2) doesn't
matter, and this is true of many systems which either have no interaction with
the outside world (e.g., deterministic CAs), or whose only inputs are
zero-memory noise (e.g., stochastic CAs, or laboratory chemical pattern formers
subject to thermal noise).  In the case of systems subject to structured input,
clearly one would like to divide increases in complexity into a portion due to
the input and a portion due to the self-organizing dynamics of the system.
There is not yet any obvious way to do this, though the ``potential response''
methods of statistical inference\cite{Holland-on-Rubin} may help.

It is appropriate, at this point, to take a step back and consider what we are
doing.  Why should we use the light-cone construction, as opposed to any other
kind of localized predictor?  Indeed, why use localized statistics at all?  Let
us answer these in reverse order.  The use of local predictors is partly a
matter of interest --- in studying self-organization, we care deeply about
spatial structure, and so global approaches, which would treat the system's
sequence of configurations as one giant time series, simply don't tell us what
we want to know.  In part, too, the local approach makes a virtue of necessity,
because global prediction quickly becomes impractical for systems of any real
size.  The number of modes required by methods attempting global prediction,
like Karhunen-Loeve decomposition, grows extensively with system
volume\cite{Zoldi-Greenside-extensive-chaos,Zoldi-et-al-extensive-scaling}.
There is thus no {\em advantage} in terms of compression or accuracy to global
methods.

The use of light-cones for the local predictors, rather than some other shape,
is motivated partly by physical considerations, and partly the nice formal
features which follow from the shape, of which we will mention
three\cite{CRS-thesis}.
\begin{enumerate}
\item The light-cone causal states, while local statistics, do not lose any
  global predictive power.  To be precise, if we specify the causal state at
  each point in a spatial region, that array of states is itself a sufficient
  statistic for the future configuration of the region, even if the region is
  the entire lattice.
\item The light-cone states can be found by a recursive filter.  To illustrate
  what this means, consider two space-time points, $(\vec{r},t)$ and
  $(\vec{q},u)$, $u \geq t$.  The state at each point is determined by the
  configuration in its past light-cone: $\localstate(\vec{r},t) =
  \epsilon(\localpast(\vec{r},t))$, $\localstate(\vec{q},u) =
  \epsilon(\localpast(\vec{q},u))$.  The recursive-filtration property means
  that we can construct a function which will give us $\localstate(\vec{q},u)$
  as a function of $\localstate(\vec{r},t)$, plus the part of the past
  light-cone of $(\vec{q},u)$ that is not visible from $(\vec{r},t)$.  Not only
  does this greatly simplify state estimation, it opens up powerful connections
  to the theory of two-dimensional automata\cite{Two-D-Patterns}.
\item The local causal states form a Markov random field, once again allowing
  very powerful analytical techniques to be employed which would not otherwise
  be available.
\end{enumerate}
In general, if we used some other shape than the light-cones, we would not
retain any of these properties.

\section{Methods and Results}

We ran the four-color, range-one CCA on a $100 \times 100$ lattice with
wrap-around boundary conditions at four different values of the $T$ parameter.
Recall that at $T = 2$, the system strongly self-organizes into spiral waves,
that $T = 3$ is also felt to be self-organizing but not as much, and that $T =
1$ and $T = 4$ show significantly less self-organization.  All four regimes
lead to stable stationary distributions.  We therefore expect $\Cmu$ to start
at zero (reflecting the uniform random initial conditions), and then rise to a
steady value which it maintains forever.  The long-run complexity should be
highest for $T = 2$, lower for $T = 3$, and much smaller for $T = 1$ or $T =
4$.

For our present purposes, where we are just interested in $\Cmu(t)$, it is
fairly straightforward to devise an algorithm to reconstruct the local causal
states from data.  At each time $t$, we determine which past and future
light-cone configurations are actually observed in the data.  Then, for each
past configuration $\localpast$, we estimate
$\Prob_t(\LocalFuture|\localpast)$, treated simply as a multinomial
distribution.  We then cluster past configurations based on the similarity of
their conditional distributions.  We cannot expect that the estimated
distributions will match exactly, so we employ a simple $\chi^2$ test (with
$\alpha = 0.05$) to determine whether the discrepancy between estimated
distributions is significant.  These clusters are then the estimated local
causal states.  We consider each cluster to have a conditional distribution of
its own, equal to the weighted mean of the distributions of the pasts it
contains.  Finally, we obtain the probabilities of the different states $s_i$
from those of their constituent past configurations, and so calculate
$-\sum_i{\Prob(s_i)\log_2{\Prob(s_i)}}$ = $\Cmu(t)$.

As a practical matter, we need to impose a limit on how far back into the past,
or forward into the future, the light-cones are allowed to extend --- their
depth.  Also, clustering cannot be done on the basis of a true equivalence
relation.  Instead, we list the past configurations
$\left\{\localpast_i\right\}$ in some arbitrary order.  We then create a
cluster which contains the first past, $\localpast_1$.  For each later past,
say $\localpast_i$, we go down the list of existing clusters and check whether
$\Prob_t(\LocalFuture|\localpast_i)$ differs significantly from each cluster's
distribution.  If there is no difference, we add $\localpast_i$ to the first
matching cluster and update the latter's distribution.  If $\localpast_i$ does
not match any existing cluster, we make a new cluster for $\localpast_i$.  (See
Figure \ref{algorithm} for pseudo-code.)  As we give this procedure more and
more data, it converges in probability on the correct set of causal states,
independent of the order in which we list past light-cones\cite{CRS-thesis}.
For finite data, the order of presentation matters, but we finesse this by
randomizing the order.

\begin{figure}[t]
\begin{tabbing}
\texttt{U} $\leftarrow$ list of all pasts in random order\\
Move the first past in \texttt{U} to a new state\\
for \= each \texttt{past} in \texttt{U}\\
\> \texttt{noMatch} $\leftarrow$ TRUE\\
\> \texttt{state} $\leftarrow$ first state on the list of states\\
\> while \= (\texttt{noMatch} and more states to check)\\
\> \> \texttt{noMatch} $\leftarrow$ (Significant difference between \texttt{past} and \texttt{state}?)\\
\> \> if \= (\texttt{noMatch})\\
\> \> \> \texttt{state} $\leftarrow$ next state on the list\\
\> \> else \=\\
\> \> \> Move \texttt{past} from \texttt{U} to \texttt{state}\\
\> \> \> \texttt{noMatch} $\leftarrow$ FALSE\\
\> if \= (\texttt{noMatch})\\
\> \> make a new state and move \texttt{past} into it from \texttt{U}
\end{tabbing}
\caption{\label{algorithm} Algorithm for grouping past light-cones into estimated states}
\end{figure}

\begin{figure}
\resizebox{\textwidth}{!}{\includegraphics{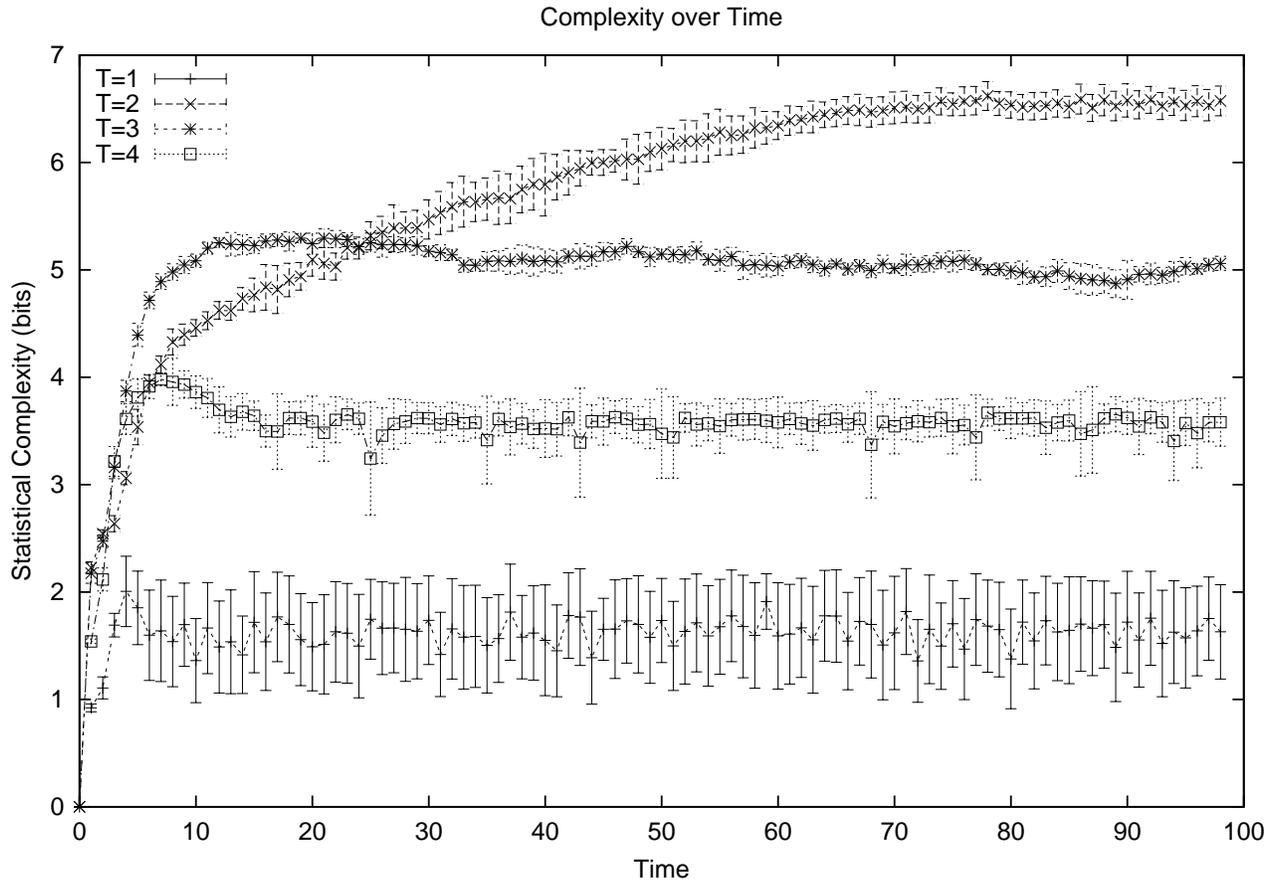}}
\caption{\label{fig:results} Complexity over time for CCA with different
  thresholds $T$. Error bars represent one standard deviation in the
  complexity; fluctuations are due to sampling effects, as described in the
  text.}
\end{figure}

Our actual numerical results, averaged over 10 independent simulations at each
value of $T$, are shown in Figure \ref{fig:results}.  They are clearly in
agreement with what we would expect if $\Cmu$ actually does measure
self-organization.  All four curves climb monotonically to steady plateaus,
leveling off when the CA configurations become stationary.  The small
fluctuations in the complexity in the stationary regimes are due to sampling
effects --- the lack of exact transitivity due to finite data inaccuracies in
the estimated conditional distributions.

\section{Conclusions}

\subsection{Directions for Future Work}

Clearly, this is just a first step towards establishing an acceptable criterion
for self-organization.  At the very least these results should be replicated
for other cellular automata, such as pattern-forming lattice
gases\cite{Rothman-Zaleski-text} and voter
models\cite{Liggett-particle-systems}.  Beyond cellular automata, our methods
are applicable, with minor modifications, to all kinds of discrete random
fields, including fields on arbitrary graphs, such as the recently-popular
``complex networks'' \cite{MEJN-on-network-structure-and-function}.  On an
irregular graph, the shape of the light-cones will generally vary from node to
node, so each node will have its own set of causal states, but it will still be
possible to calculate $\Cmu$ over the graph, and so apply the basic ideas set
out here.

Another obvious extension is to work with experimental data.  Digital video
provides discrete-valued data on regular, discrete lattices, and as such is
eminently suited to our approach.  Ultimately, one would like to be able to
predict when, and how much, different physical systems will self-organize and
actually validate those predictions with experimental data.  While there are
important issues in choosing appropriate discretizations \cite{Badii-Politi},
there does not seem to be any obstacle, in principle, to doing so.

\subsection{Summary}

It would be nice to have a general theory of self-organization.  Before work on
that theory can begin, we need a mathematical characterization of
self-organization, preferably one which can be applied to data directly.
``Decline in thermodynamic entropy'' is not really suited for this role, but
``increase in complexity'' is.  We have shown how to define a sensible
complexity measure for spatio-temporal systems, based on the amount of
information actually required for optimal statistical prediction.  This
statistical complexity, $\Cmu$, can be reliably and straight-forwardly
estimated from data.  In the particular case of cyclic cellular automata, where
the qualitative behavior changes drastically with the threshold parameter $T$,
we find that whether or not $\Cmu$ increases over time agrees completely with
intuitive judgments about self-organization.

\acknowledgments

CRS's work was supported by a grant from the James S. McDonnell Foundation.  We
thank Derek Abbott, Dave Feldman, Janko Gravner, David Griffeath, Rob
Haslinger, Cris Moore, Scott Page, Eric Smith and Jacob Usinowicz for valuable
conversations, and Kara Kedi for moral support and last-minute typing.

\bibliography{locusts} \bibliographystyle{spiebib}
\end{document}